\documentclass[11pt]{article}
\usepackage{amsmath}
\usepackage{graphicx}
\def\RG{{\it RG\/}}
\def\SM{{\it SM\/}}
\def\MSSM{{\it MSSM\/}}

\def\ikap{\kappa^{-1}}
\def\iikap{\kappa^{-2}}
\newcommand{\msbar}{$\overline{\mbox{\small MS}}$}
\def\st{\scriptstyle}

\def\be{\begin{equation}}
\def\ee{\end{equation}}
\def\bea{\begin{eqnarray}}
\def\eea{\end{eqnarray}}
\def\frak#1#2{{\textstyle{\frac{#1}{#2}}}}
\def\GeV{\hbox{GeV}}
\def\TeV{\hbox{TeV}}
\def\vev#1{\mathopen\langle #1\mathclose\rangle }
\def\nn{\nonumber\\}
\def\pa{\partial}
\def\ga{\gamma}  
\def\zbar{{\overline{z}}}
\def\psib{\overline{\psi}}
\def\betab{\overline{\beta}}
\def\gab{\overline{\gamma}}
\def\lab{\overline{\lambda}}
\def\xib{\overline{\xi}}
\def\eff{\st eff}

\begin{document}
\begin{titlepage}
\begin{flushright}
LTH 738\\
NSF-KITP-07-13\\
\end{flushright}

\vspace*{3mm}

\begin{center}
{\Huge The Effective Potential, the Renormalisation Group  
and Vacuum Stability}\\[12mm]

{\bf Martin B.~Einhorn}\footnote{meinhorn@kitp.ucsb.edu}\\

\vspace{5mm}
Kavli Institute for Theoretical Physics, 
University of California, Santa Barbara
CA 93106-4030, USA.\\

\vspace{3mm}

{\bf and  D.~R.~Timothy~Jones}\footnote{drtj@liverpool.ac.uk}\\

\vspace{5mm} 
Dept. of Mathematical Sciences,
University of Liverpool, Liverpool L69 3BX, UK.\\
\vspace{8mm}
\end{center}

\begin{abstract} 
We review the calculation of the the effective potential with 
particular emphasis on cases when the tree potential or the 
renormalisation-group-improved, radiatively corrected potential exhibits
non-convex behaviour.  We illustrate this in a simple Yukawa model which
exhibits a novel kind of dimensional transmutation.  We also review briefly  
earlier work on the Standard Model.  We conclude that, despite some
recent claims to the contrary, it can be possible to infer reliably that
the tree vacuum does not represent the true ground state of the theory. 
\end{abstract}

\vfill
\end{titlepage}

\section{Introduction}

The effective potential $V(\phi)$ has proved to be an invaluable tool 
for investigating the nature of the vacuum state in weakly coupled 
quantum field theories. There is a well defined
prescription~\cite{cw},\cite{jackiw}\  facilitating the perturbative
calculation of $V(\phi)$, and a large  literature of such calculations
for various field theories including the  Standard Model
({\it SM})~\cite{fjj}\ and the Minimal Supersymmetric Standard Model
(\MSSM)~\cite{spmartin}.  
Theoretical forays into the 
Early Universe frequently involve the effective potential and, 
for example, transition from an unstable or metastable state by tunnelling 
or  ``slow-roll".\cite{Boyanovsky:2006bf},\cite{Quiros:1999jp}.
New scalar excitations such as the inflaton
are hypothesised to lead to the desired behaviour.\cite{Kolb:1994eu}. 

In this article, we confine ourselves largely to one specific topic: 
the destabilisation of a tree-level vacuum by radiative corrections.
In the {\it SM}, it has been argued that, if the fermion loop contribution 
(dominated by the top quark) to
the  one-loop potential were large enough then the electroweak vacuum would be
destabilised,  due to the existence of a deeper minimum at large $\phi$;
this phenomenon is translated into a lower limit on the Higgs mass as  a
function of the top quark mass~\cite{Hung:1979dn}-\cite{isido}. 

The electroweak minimum of the \SM\ is a consequence of the choice 
$m^2 < 0$ for the  Higgs mass parameter. As a result the perturbative
definition of $V(\phi)$  is non-convex (in the tree approximation) and
develops an imaginary part  for small values of $\phi$ at one loop and
beyond.  These issues  were addressed in the well-known work of Weinberg
and Wu~\cite{wewu}, where it was shown that  the imaginary part has a
natural interpretation as a decay rate of a well-defined state,  and
that the perturbative $V$, although non-convex, is nevertheless a 
physically meaningful quantity.  In the \SM\ both issues (imaginary part
and  non-convexity) arise once again in the neighbourhood of the
fermion-induced instability,  but it has generally been believed that
the perturbative $V(\phi)$ remains  meaningful in this region too. This point 
of view has, however, been challenged recently~\cite{BF}-\cite{holland}, which 
provides us with one motive for this paper. We devote our attention chiefly 
to the basic $\lambda \phi^4$ theory and a simple extension involving a 
Yukawa coupling; the latter exhibits some interesting features including 
a form of dimensional transmutation reminiscent of but distinct from the 
case discussed originally by Coleman and Weinberg (CW) in Ref.~\cite{cw}.

\section{The Effective Potential}\label{sec:effpot}

To introduce some notation, we review a few definitions.  The generating
function $W[{\cal J}(x)]$ for connected Green's functions may be defined
from the (Euclidean) Feynman path integral
\be
Z[{\cal J}]\equiv e^{-W[ {\cal J}]}=\int {\cal D}\Phi~e^{-S[\Phi]-\int dx {\cal J}(x)\Phi(x)},
\ee
where $S$ is the classical action and $\Phi$ is represents all fields of the theory (with indices suppressed.)  The classical field associated with the source is 
\be\label{phiclass}
\Phi_{\cal J}\equiv \frac{\delta W}{\delta {\cal J}(x)},
\ee
and the effective action is defined via the Legendre transform
\be
\Gamma_{eff}[\Phi_{\cal J}]=W[ {\cal J}]-\int dx {\cal J}(x)\Phi_{\cal J}(x).
\ee
The converse of Eq.~(\ref{phiclass}) is
\be
{\cal J}(x)= - \frac{\delta \Gamma_{eff}}{\delta \Phi_{\cal J}}.
\ee

The effective potential is the response to a constant source ${\cal J}(x)\equiv j.$  Assuming that the associated classical field is also constant, $\Phi(x)\equiv\phi_j,$ the effective action functional becomes the effective potential
\be\label{veff}
\Gamma_{eff}[\Phi_{\cal J}]=V_{\eff}(\phi)\int dx,
\ee
with
\be
j= - \frac{\partial V_{\eff}}{\partial \phi}.
\ee

There are some significant qualifications of this formalism that must be
kept in mind.  The mapping between the source and the classical field
may be multi-valued, so different branches of solutions must be
discussed.   Moreover, it is well-known~\cite{Iliopoulos:1974ur}\  that the
exact effective potential is convex and that the true effective
potential looks roughly like the convex hull of the classical effective
potential.  Between classically  degenerate minima, the system may break
up into domains having one value of the field or the other (Maxwell's
construction.)  Thus, the assumption that a constant source is
associated with a spacetime  independent classical field breaks down.

Another point of view~\cite{coleman} is to consider the theory
associated with the classical potential
\be\label{jtheory}
U_j(\phi)\equiv V_{cl}(\phi) + j \phi,
\ee
where $j$ is thought of as a coupling constant, chosen so that the
expectation value of the field, defined (at the tree level) by the equation
\be\label{eq:umin}
\frac{\pa U_j}{\pa\phi} = 0
\ee
is a particular value\footnote{Since 
$j$ has dimensions of mass-cubed, the $\beta$-functions for
other couplings are unchanged (in a mass-independent renormalization
scheme.)} $\phi_j.$   The effective potential is the expectation value of the
Hamiltonian density for this modified theory.  Quite a lot can be
learned by thinking in terms of $U_j(\phi),$ and we will return to this
below.


With some normalization convention, the effective potential may be
summarised by the equation~\cite{jackiw}\
\be\label{effjackiw}
V_{\eff}(\phi)=V_{cl}(\phi)+\frac{1}{64\pi^2}
{\rm STr}~M(\phi)^4\ln\left(\frac{M(\phi)^2}{\mu^2} \right) 
-{\cal W}_2(\phi).
\ee
In this expression, $\phi$ is the background field; $V_{cl}$ is the
classical or tree potential; $M(\phi)^2$ is the mass matrix in tree
approximation associated with the various particles in the theory and 
$\rm STr$ is the ``supertrace," a sum over bosonic and fermionic degrees of
freedom.  The supertrace term comes from the one-loop approximation,
while  ${\cal W}_2$ represents the two-loop and higher contributions and
non-perturbative contributions.  It is given by  \be {\cal W}_2(\phi)
\equiv \ln\left<\exp\left(-\int d^4x{\cal
L}_I(\phi,\widetilde\phi)\right)\right>. \ee Here $\widetilde\phi$ is
the quantum field defined by replacing $\phi\to\phi+\widetilde{\phi}$ in
the original Lagrangian.  The precise definition of the interaction
Lagrangian ${\cal L}_I$ and the vacuum expectation value are given in
Ref.~\cite{jackiw}; since they play no role in this paper, we will not pause
to define them here.  The expression for $V_{\eff}$ in
Eq.~(\ref{effjackiw}) is {\bf formally} exact, but most useful for
generating the loop expansion.  

\subsection{The $\phi^4$ Model}\label{subsec:phi4}

In the next section, we will discuss a Yukawa model.  First we review
the well-known, real scalar $\phi^4$-model, with 
\be
V_{cl}=\frac{m^2}{2} \phi^2+\frac{\lambda}{4!}\phi^4
\ee
so $M^2(\phi)=V_{cl}''(\phi)=m^2+\lambda\phi^2/2,$ and the one-loop
formula becomes 

\be\label{oneloop}
V_1=V_{cl}(\phi)+\frac{V_{cl}''(\phi)^2}{64\pi^2} 
\ln\left(\frac{V_{cl}''(\phi)}{\mu^2}\right).
\ee
We will discuss three cases depending on the value of $m^2.$  For the
$\phi^4$-model with $m^2>0,$ the coupling $\lambda$ is infra-red (IR)
free.  As  $\phi\to 0,$ there is no problem with the perturbative
expansion, and the origin remains a minimum.  No large logarithms arise
so long as the normalization scale $\mu\sim O(m).$  As $\phi\to\infty$,
the one-loop correction becomes large, so that one must improve on this
perturbation expansion.  Loosely speaking, one must take $\mu^2 \sim
V_{cl}''(\phi).$  However, since $\lambda$ becomes large at large
scales, perturbation breaks down in any event.  Whether the model has a
sensible strong coupling solution is not known analytically.  Lattice
calculations \cite{Luscher:1988gc}-\cite{Callaway:1988ya}  
strongly suggest that in fact the theory
requires a cutoff (i.e., is trivial in the continuum limit.)   

For $m^2=0,$ the logarithm diverges as $\phi\to0,$ and the effective
potential must be renormalization-group-improved to remove such large
logarithms~\cite{cw}.  The conclusions turned out to be  the same
as in the case $m^2>0.$

For $m^2<0,$ the case of spontaneous symmetry breakdown, the situation
is more complicated.  The double-well potential (Fig.~1) has two
degenerate minima.  

\begin{figure}
\begin{center}
\includegraphics[angle = 0,totalheight=8cm]{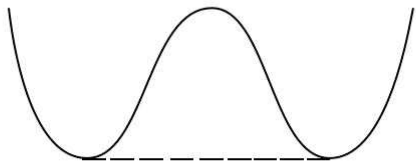}%
\end{center}
\caption{The Double well potential}
\end{figure}

In perturbation theory, the logarithm diverges where
$V_{cl}''(\phi)=0,$ while for $V_{cl}''(\phi)<0,$ the effective
potential becomes complex.  The  imaginary part is
well-known~\cite{coleman}\ to be related to the decay rate per unit
volume for the state of the field.  Therefore, such an effective
potential, based on the assumption of a stable value of the field,
 can only be valid for a limited time.  Its precise definition 
 therefore requires some care and has been discussed in detail by
Weinberg \& Wu~\cite{wewu}.  The preceding formalism is insufficient,
since the exact effective potential is always real and
convex~\cite{Iliopoulos:1974ur}.   Between the two minima, the true
effective potential is expected to look like the dashed line in Fig.~1. 
 The crucial requirement to give meaning to results such as
Eq.~(\ref{oneloop}) is that, for some period of time short compared to
the decay time, the classical background state is required to be homogeneous.  
This is a metastable situation which becomes invalid after sufficiently long time.
The true ground state for this range of values of the field is
described by a breakup into domains within which there are approximately
constant but different values of the background.   This approximation is
even experimentally meaningful in certain situations such as super-cooled
steam or super-heated water.  

Substantial insight into what happens for small coupling $\lambda$ may
be obtained by considering the theory associated with
Eq.~(\ref{jtheory}).  For small $j>0,$ the potential looks as in Fig.~2.

\begin{figure}\begin{center}
\includegraphics[angle = 0,totalheight=8cm]{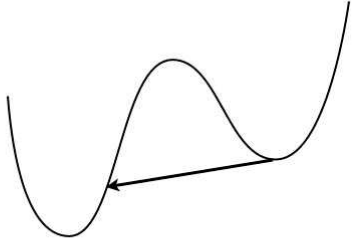}%
\end{center}
\caption{The Double well potential for small $j$.}
\end{figure}

Classically there are 3 solutions for $\phi_j$ to 
Eq.~(\ref{eq:umin});  $U_j$ still has two
local minima but they are no longer degenerate.  The one at $\phi_j>0$
is metastable, decaying eventually non-perturbatively via tunnelling. 
Thus, the effective potential for this homogeneous background acquires a
nonperturbative imaginary part.   The third solution is a local maximum
at a value of $\phi_j$ near zero in the perturbatively
unstable region $V_{cl}^{''}<0.$  As a result, the one-loop approximation
Eq.~(\ref{oneloop}) has an
imaginary part.  This instability for long wavelength fluctuations has
been discussed in some detail by Guth \& Pi~\cite{Guth} and expanded on in
Ref.~\cite{wewu}.  The breakdown of the homogeneous state is described
therein, and we have little to add.\footnote{One must be careful about
the interpretation of the ``long-time" regime as described in Ref.~\cite{Guth}. 
That is only an intermediate time at best; eventually the system
descends to the mixed state described by the convex effective potential.}

As $j>0$ increases, the metastable minimum approaches the local maximum,
and eventually they merge, at which point $U_j^{''}(\phi_j)=0.$  This is
depicted in Fig.~3.  

\begin{figure}
\begin{center}
\includegraphics[angle = 0,totalheight=8cm]{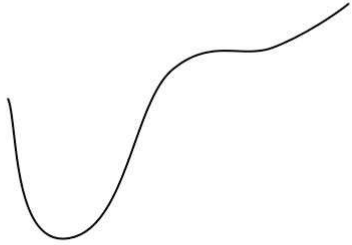}%
\end{center}
\caption{The Double well potential when $U_j^{''}(\phi_j)=0.$}
\end{figure}

Since $U_j^{''}=V_{cl}^{''},$ this is also the point
where $M^2(\phi)=0.$  To our knowledge, the behaviour 
of the system at this particular value of $\phi_j$ 
has not been analysed in the literature.
Even though the imaginary part of the one-loop effective
potential vanishes at this point, it is perturbatively unstable, since, as noted
in Ref.~\cite{efjs}, the effective potential has {\it power\/} divergences at
this point in orders higher than three loops.  
Specifically, elementary power counting demonstrates that 
at $L$ loops, $V(\phi)$ includes contributions of the form 
$\kappa^3 (\lambda\phi)^4\eta^{L-3}$, where
\be
\eta = \frac{\kappa\lambda^2\phi^2}{M^2}.
\ee
where $\kappa = (16\pi^2)^{-1}$.
This phenomenon does  not appear to have been discussed in the
literature, presumably because it  begins at four loops.  It should be
no surprise, however,  given the situation depicted in Fig.~3.  This is an
inflection point where fluctuations of one sign do not grow, but
fluctuations of opposite sign run away.  Given this classical
instability, one  should not be surprised to find a divergence  in
perturbation theory.  One would expect the growth of perturbations to be
less dramatic  than in the region of negative curvature.  Whereas the
correlation length for unstable modes in the regime where $M^2(\phi)<0$
is finite, for $M^2(\phi)=0$ it is infinite. However, there is a mass
scale in the classical potential, viz., 
$V_{cl}^{'''}(\phi_j)=\lambda\phi_j=\sqrt{2\lambda |m^2|}\equiv\Delta,$
and one would expect this to determine the scale of growth of perturbations.

Precisely how these evolve in time has not been determined, but
presumably a treatment similar to Ref.~\cite{Guth}\ should be possible.  We
have not carried such an analysis, but if, after some time, the system
evolves essentially classically, then it is easy to work out the
behaviour.  For the zero mode, it is clear from the figure that, if the
field starts anywhere other than at the inflection point with zero
velocity, it will oscillate in time.  Regardless of where it is
initially, it eventually will roll down the hill, far from the
starting point, indicating the instability of small displacements.  For
short times, one can show that the displacement grows as $t^2,$ with a
coefficient proportional to $\Delta.$  For the quantum mechanical
problem, one would have to develop a probability distribution of the
displacement similar to Ref.~\cite{Guth}.  

\section{The Yukawa model}

Here we consider a simplified model which omits gauge fields; it is also
the model used as an introductory  example in Ref.~\cite{BF}.  The model
resembles the \SM\ in that it displays the phenomenon of vacuum
instability for  sufficiently large Yukawa coupling, while, as we shall
emphasise,  differing in some crucial respects. It has the advantage
that  explicit solutions to the (one-loop) renormalisation group (RG) 
equations are easily constructed, so the RG evolution is particularly
transparent. In this section we will analyse the renormalisation group 
evolution of the mass and couplings and consider the nature of the theory 
at different scales, while in the next section we will 
consider in detail the scalar effective potential.

The model consists of a real scalar field coupled to a 
set of $n_F$ massless Dirac fermions;
the Lagrangian  is 

\be
L = \frak{1}{2}\pa^{\mu}\phi\pa_{\mu}\phi +i\psib_i\ga^{\mu}\pa_{\mu}\psi_i 
-\frak{1}{2}m^2\phi^2-\frak{1}{24}\lambda\phi^4 -h\phi\psib_i\psi_i.
\ee

In this section we will mainly assume $m^2 > 0$, and make a few comments about 
the cases $m^2 < 0$ and $m^2 =0$ at its end; we will return to these 
cases in more detail in Section~4. 

The one-loop $\beta$-functions for the mass $m^2$ and the couplings 
$h$ and $\lambda$ are given by
\bea
\beta_{m^2} &=& \kappa m^2\left(\lambda + 4 n_F h^2\right)\nn
\beta_{\lambda} &=& \kappa\left(3\lambda^2+8n_F \lambda h^2 -48n_F h^4\right)\nn
\beta_{h} &=& (2n_F + 3) \kappa h^3,
\eea
where $\kappa = (16\pi^2)^{-1}$. We will also require the scalar anomalous dimension, 
which at one loop is 
\be
\gamma_{\phi} = 2\kappa n_F h^2.
\ee
To analyse the RG behaviour of the model it is convenient to consider 
$Y=\lambda / h^2$, which satisfies
\be
\beta_Y = \kappa h^2\left(3Y^2 +(4n_F-6)Y -48n_F\right),
\label{eq:betaY}
\ee
so that for $n_F = 1$,  $Y$  
has fixed points $Y_{+,-} = \frac{1}{3}(1\pm\sqrt{145})
\approx 4.35, -3.68$.  In Fig.~4, we plot $\beta_Y$ against $Y$ for this case.

For all $n_F \geq 0$, $Y_+ > 0$ and $Y_{-} < 0$; 
and for large $n_F$,\footnote{Of course 
if we simply increase $n_F$ the theory soon 
loses a perturbative regime. 
We will consider separately the large $n_F$ theory 
obtained by  $h \to h/ \sqrt{n_F}$}  $Y_{+,-} \approx 12, -10-4n_F/3.$
\begin{figure}
\begin{center}
\includegraphics[angle = 0,totalheight=10cm]{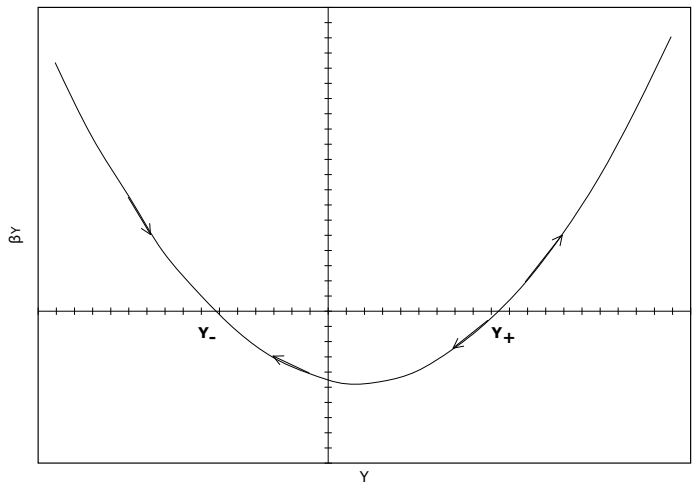}%
\end{center}
\caption{Plot of $\beta_Y$ against Y (for $n_F = 1$). 
The arrows show the direction of RG 
flow as $t$ increases.}
\end{figure}
It is easy to see that $Y_{+,-}$ are IR and UV (ultra-violet) 
attractive fixed points
respectively.  Let us consider the evolution of the couplings, 
starting at $t= \ln\mu = 0$; and first look at what  happens in the UV, i.e. for
$t > 0$. The evolution  of $h$ is elementary:
\be
h^2 (t) = \frac{h_0^2}{1-B\kappa h_0^2 t}.
\ee
where $B =2(2n_F+3)$. As $t$ increases, $h(t)$ approaches a Landau pole,
thus evidently departing the  perturbatively believable regime. If $Y(0)
> Y_+$, then $Y(t)$ increases with $t$; eventually $h$ would reach its
Landau pole,  though in fact, in one-loop approximation, 
$\lambda$ reaches its Landau pole first.  To  see this, we require the explicit
solution for $Y$ of  Eq.~(\ref{eq:betaY}), which for $Y(0) \equiv Y_0 >
Y_+$  is 
\be
Y(t) = \frac{Y_+ - z(t)Y_{-}}{1-z(t)}
\ee
where 
\be
z(t) = \frac{Y_0-Y_+}{Y_0-Y_{-}}
\left(\frac{h^2}{h_0^2}\right)^{\frac{3(Y_+-Y_{-})}{B}}
\label{eq:ysol}
\ee
and we see that $Y\to\infty$ when $z(t)\to 1$, which manifestly occurs 
for a finite value of $h(t)$. 

If $ Y_{-} < Y_0  < Y_+$ then it is clear from Eq.~(\ref{eq:betaY}) that  as
$t\to \infty$ we will have $Y(t) \to Y_{-}$. Moreover at some finite value
of $t$, $t^{\prime}$ say, $Y(t)$ and hence $\lambda(t)$ passes through zero.
Clearly the issue of perturbative  believability of this result rests
simply on the value of $h(t^{\prime})$.   The explicit solution for $Y$ in this region 
is given by 
\be
Y(t) = \frac{Y_+ \zbar(t) + Y_{-}}{1+\zbar(t)}
\ee
where 
\be
\zbar(t) = \frac{Y_0-Y_{-}}{Y_+ -Y_0}
\left(\frac{h_0^2}{h^2}\right)^{\frac{3(Y_{+}-Y_{-})}{B}}
\ee
so that at $t = t^{\prime}$ we have 
\be
\frac{h^2}{h_0^2} = 
\left[\frac{Y_{+}(Y_0-Y_{-})}{Y_{-}(Y_0-Y_{+})}\right]^\frac{B}{3(Y_{+}-Y_{-})}
\ee
For $t > t^{\prime}$, $Y \to Y_{-}$; and since $h(t)$ increases with $t$ 
then so does the magnitude of  $\lambda$. Thus the scalar 
potential becomes negative;
and according to the one-loop solution  unbounded from below.

Finally for $Y < Y_{-}$, the solution for $Y$ is again  given by 
Eq.~(\ref{eq:ysol}) and as $t$ increases $Y$ increases towards the fixed
point $Y_{-}$.

Now let us discuss what happens in the infra-red, i.e. as $t$ decreases
from zero. Let  us assume that the scale $\mu_0$ corresponding to $t=0$
satisfies $\mu_0^2 > m^2$.   Since $Y_{+}$ is IR attractive, $Y$ will
approach $Y_{+}$ both for $Y_0 > Y_{+}$ and $Y_{-} <  Y_0 < Y_{+}$,  while if
$Y_0 < Y_{-}$, $Y(t)$ decreases as $t$ decreases. The situation changes,
however,  when $\mu^2$ reaches $m^2$. 

It is well-known that a mass-independent renormalization scheme, such
as \msbar, does not properly take into account threshold effects, so
one must be careful extrapolating to scales $\mu^2 < m^2.$   To obtain 
the correct form of the effective field theory we must integrate out the
massive scalar excitations (just as  we do not include the top quark
contribution to the QCD $\beta$-function in describing physics  at
scales of a few GeV, for example). The result is a non-renormalisable
theory of massless fermions with  a series of interactions suppressed by
powers of $1/m^2.$   One may equally well retain the scalar field as an
auxiliary field.  In the language of Feynman diagrams, we rewrite each
scalar propagator as  \be \frac{1}{p^2+m^2} = \frac{1}{m^2} \cdot
\frac{1}{1+p^2/m^2}. \ee The factor of $1/(1+p^2/m^2)$ is then expanded
as a power series in $p^2/m^2.$   In this way, the scalar field becomes
an auxiliary, i.e.,  non-propagating, field.   To organise the effective
field theory, it is useful to associate the factor of $1/m^2$ with the
vertices at each end of the propagator and the factors of $p^2$ with
derivatives on the fields.  Correspondingly, we rescale 
$\sigma \equiv m\phi,$ and thus are led to the Lagrangian  
\be L =
i\psib_i\ga^{\mu}\pa_{\mu}\psi_i  +\frak{1}{2}\sigma^2- \frac{h}{m}
\sigma \psib_i\psi_i +\frac{\alpha_1}{m^2}(\partial_\mu\sigma)^2  
-\frac{1}{24}\lambda(\frac{\sigma}{m})^4  -\frac{\alpha_2}{m^4}
(\psib\partial_\mu \psi)^2 + \cdots
\label{eq:IRlimit} 
\ee 
All terms but the last are just a rewriting of the original Lagrangian,  so it appears
to be a sleight-of-hand.\footnote{Actually, in writing
Eq.~(\ref{eq:IRlimit}), we have dropped a number of redundant operators
using the lowest order equations of motion.  For the justification, see
ref.~\cite{einwud}.}   However, now the ``kinetic energy" term (with
coefficient $\alpha_1$) is to be thought of as a higher-dimensional
operator.  To leading order, $\alpha_1 \sim O(h^2).$  (Note that
$\sigma$ has dimensions of $(\hbox{mass})^2$, like an auxiliary field of
a supersymmetric theory.)   To lowest order in $1/m,$ the auxiliary
field is in fact given by 
\be \sigma = \frac{h}{m}\psib\psi. 
\ee
It is easy to see from Eq.~(\ref{eq:IRlimit}) that the $\sigma^2$ term  
and the Yukawa term are unrenormalised, while the fermion box diagram with 
four Yukawa insertions gives rise to a $\beta$-function 
for $\lambda$ of the form 
\be
\beta_{\lambda} = -48\kappa n_F h^4,
\label{eq:blamir}\ee
so that 
\be
\lambda(t) = \lambda(0) -48\kappa n_F h^4 t. 
\label{eq:lamir}\ee
Here we take $t\equiv \ln(\mu/m),$ so that $t<0$ in the low-energy
theory.  The value of  $\lambda(0)$ is obtained by matching to the
corresponding coupling in the original theory on scales $\mu> m.$  
Moreover, in the expansion in $1/m,$  Eq.~(\ref{eq:blamir}) and
Eq.~(\ref{eq:lamir})  are exact. 

Note that $\lambda$ grows in the IR limit.  Now by assumption,
$\lambda(0)>0,$ so that the tree vacuum is at $\phi = 0$.
(Otherwise, the fermions would be massive and the low-energy effective
field theory would be different.)  Thus the self-interaction of the
scalar is repulsive and remains so for all scales in the low-energy
theory.  Intuitively at least, 
we would not expect spontaneous symmetry breaking in this regime, 
a conclusion that will be reinforced by the discussion in the next section.

Let us next consider the case classically scale-invariant case $m^2 = 0$. 
Let us suppose that, at  $t=0,$ we have $Y(0) > Y_{-}$. Then in the IR 
(since the scalar field does not now decouple), 
$Y$  flows  to the fixed point $Y = Y_{+}$. 
Since $h \to 0$ in the IR, we therefore have also $\lambda \to 0$. Thus 
the scalar potential tends to zero smoothly and the origin ($\phi = 0$)
remains a local minimum of $V(\phi)$.

In the UV, the RG behaviour of the  couplings is exactly as in the $m^2
> 0$ case; thus  for  $Y_{-} < Y(0) < Y_+$, $\lambda$ becomes negative,
and the potential unbounded  from below. As we shall see in the next
section, when we consider  the radiatively corrected potential, the
theory actually undergoes  dimensional transmutation\cite{cw} leading
to interesting structure of the effective potential. Unlike the scalar
QED case, the dimensional quantity manifested is in the field value at a
{\it maximum\/} of the potential where it turns over.  We will consider
this case (and the $m^2 < 0$ case) in more detail in the next section.  

\subsection{The large $n_F$ limit}

In order to define a weakly-coupled large $n_F$ theory 
we must let $h \to h/\sqrt{n_F}$. 
Then we have for the $\beta$-functions and $\gamma$ the following expressions.
For $\mu^2 > m^2$:
\bea
\beta_{m^2} &=& \kappa m^2\left(\lambda + 4 h^2\right)\nn
\beta_{\lambda} &=& \kappa\left(3\lambda^2+8\lambda h^2\right)\nn
\beta_{h} &=& 2\kappa h^3\nn
\gamma_{\phi} &=& 2\kappa h^2.
\label{largen}
\eea

With $Y=\lambda / h^2$ once again, we have from Eq.~(\ref{largen})
\be
\beta_Y = \kappa h^2\left(3Y^2 +4Y\right),
\label{eq:betaYn}
\ee
so that in the UV the analysis remains similar, but with   
points $Y_{+,-} = 0, -4/3$.

In the IR limit we see that (due to the rescaling $h \to h/\sqrt{n_F}$)
we have $\beta_{\lambda} = 0,$ so that for $\mu < m,$ the coupling 
$\lambda$ freezes, remaining constant for $t \to -\infty$.  

In the next section we explore the consequences of these RG considerations 
for the effective potential. 

\section{The RG improved potential}\label{sec:RGI}

Ref.~\cite{efjs}\ contains a detailed analysis of the procedure 
for RG-improving the effective potential; and in particular the 
significance of the cosmological constant term $\Omega$ (previously generally 
neglected) was emphasised. It was also remarked that by considering 
the renormalisation group improvement of 
$V^{\prime} \equiv \frac{\pa V}{\pa \phi}$  the $\Omega$-issue may be finessed. 
Here we will compare the RG improved forms of $V$ and $V^{\prime}$. 

We will here use the RG improved forms as follows:  
\bea 
V(\mu, \lambda_i, \phi) &=& \xib(t)^4 V 
(\mu(t), \lab_i (t), \phi)\nn
V^{\prime}(\mu, \lambda_i, \phi) &=& \xib(t)^4 V^{\prime} 
(\mu(t), \lab_i (t), \phi),
\eea
where
\bea
\frac{ d\lab_i (t)}{dt} &=& \betab_i\left(\lab(t)\right)\nn
\betab_i &=& \frac{\beta_i + \delta_i\lambda_i \gamma}{1+\gamma}\nn
\gab_i &=& \frac{\gamma}{1+\gamma}\nn
\xib(t) &=& \exp\left(-\int_0^t\,
\gab\left(\lambda_i (t^{\prime})\right)\, dt^{\prime}\right)\nn
\mu(t) &=& \mu e^{t}.
\eea
Here $\lambda_i$ stands for all couplings and masses, with canonical mass 
dimension $\delta_i$ and $\beta_i$ is the $\beta$-function for $\lambda_i$.
Evidently to leading order we have $\gab = \gamma$. 
The effective potential $V(\phi)$ is given (in the 
one-loop approximation) as follows:
\bea\label{effpot}
V(\phi) &=& \Omega(\mu, m, \lambda)  + \frak{1}{2}m^2\phi^2 + 
\frak{1}{24}\lambda\phi^4\nn
&+& \frac{1}{64\pi^2}\left[M^4
\left(\ln\frac{M^2}{\mu^2}
-\frac{3}{2}\right)
-4n_F h^4\phi^4\left(\ln\left(\frac{h^2\phi^2}{\mu^2}\right)
-\frac{3}{2}\right)\right]
\eea
where $M^2 = m^2+\frak{1}{2}\lambda\phi^2$, and $\Omega$ is the afore-mentioned 
cosmological constant term (this was denoted $\Omega^{\prime}$ 
in Ref.~\cite{efjs}).  

Our RG improved solutions thus take the form:
\bea
V(\phi) &=& \xib(t)^4\biggl[\Omega(t)  + \frak{1}{2}m^2(t)\phi^2 + 
\frak{1}{24}\lambda(t)\phi^4
+ \frac{1}{64\pi^2} M^4(t)
\left(\ln\frac{M^2(t)}{\mu^2(t)}
-\frac{3}{2}\right)\nn
&-&\frac{1}{64\pi^2} 4n_Fh^4(t)\phi^4\left(\ln \frac{h^2(t)\phi^2}{\mu^2(t)}
-\frac{3}{2}\right)\biggr],
\label{eq:vrg}
\eea

\bea 
V^{\prime} &=& \xib(t)^4\biggl[m^2 (t) \phi + 
\frak{1}{6}\lambda(t) \phi^3
+ \frac{1}{32\pi^2}\lambda(t) M^2 (t)
\left(\ln\frac{M^2 (t)}{\mu^2 (t)}
-1\right)\nn
&-&\frac{1}{32\pi^2} 8n_Fh^4 (t) \phi^3\left(\ln\frac{h^2 (t) \phi^2}{\mu^2 (t)}
-1\right)\biggr].
\label{eq:dvrg}
\eea

In order that $V$ satisfy the usual RG equation, 
$\Omega(\mu,\lambda_i)$ must itself satisfy  an RG equation 
which to leading order can be written 
\be\label{eq:Omegarg}
\left[\mu\frac{\pa}{\pa\mu} + \beta_i\frac{\pa}{\pa\lambda_i}\right] \Omega  = 
\frac{1}{32\pi^2}m^4.
\ee
In Ref.~\cite{Kastening}\ and Ref.~\cite{efjs}\ the consequences of 
various choices for $\Omega$ were considered. For example, one may decide 
to require that $\Omega$ be free of explicit dependence on $\mu$; 
in the absence of the Yukawa coupling $h$ the 
resulting solution of Eq.~(\ref{eq:Omegarg}) (using one-loop $\beta$-functions) 
is  
\be\label{eq:Omegasol}
\Omega = -m^4 \left(\frac{1}{2\lambda} 
+ \frac{c}{\lambda^{\frac{2}{3}}}\right),
\ee
where $c$ is an arbitrary constant. It is interesting that when we 
renormalisation improve the solution as described above, the $c$-term 
in Eq.~(\ref{eq:Omegasol}) remains independent of $t$ in this case because we have
\bea
\lambda(t) &=& \frac{\lambda_0}{1-3\kappa\lambda_0 t}\nn
m^2 (t) &=& \frac{m^2 (0)}{(1-3\kappa\lambda_0 t)^{\frac{1}{3}}}.
\eea
The case $h \neq 0$ is not so easy to analyse. 

A more natural form for $\Omega$ is 
\be
\Omega = -\hat V(\phi)\Bigm\vert_{\phi=v}
\ee
where $\hat V(\phi)$ is the potential omitting $\Omega$, and 
$v$ is the value of $\phi$ at any extremum of $\hat V$.
Of course the simplest such extremum is $v = 0$, corresponding to  
a solution for $\Omega$ which is (to leading order) 
\be 
\Omega = -\frac{1}{64\pi^2}m^4\ln \frac{m^2}{\mu^2}.
\label{eq:cosm}
\ee
This corresponds of course to simply subtracting  the value of $V$ at
$\phi = 0$.  It is this form of $\Omega$ we use in the subsequent analysis. 

Let us now  consider the behaviour of the potential as a function of
$\phi$.  We will consider separately the three cases 
$m^2 > 0$, $m^2 = 0$ and $m^2 < 0$.

\subsection{The $m^2 > 0$ case}

For $\phi \to 0$,    if we choose $t$ so that $\mu(t) = \phi,$
then  this controls the logarithm  in the fermion loop contribution; the
scalar loop contribution  decouples for $\mu^2 < m^2$ (as described in
the last section for the $\beta$-functions)  and must hence be removed
from Eqs.~(\ref{eq:vrg},\ref{eq:dvrg})\,  and we replace
Eq.~(\ref{eq:cosm})\  by $\Omega = 0$.  The behaviour of the potential
for $\phi^2 < m^2$  is then determined by the IR RG evolution of 
$\lambda$ which we explored  in the last section, 
culminating in Eq.~(\ref{eq:lamir}). From this it is clear that, 
with $\mu \sim \phi$, $\lambda$ grows like $\lambda \sim - \ln \phi$, and 
hence $V \sim m^2\phi^2/2 + \lambda\phi^4/24 \to 0$.  

The
question to be addressed now is whether we can conclude that this represents
the ground state of the theory; so let us examine the potential for
large $\phi$. Once  again we choose $t$ so that $\mu = \phi$, in order
to control  the logarithms in Eqs.~(\ref{eq:vrg},\ref{eq:dvrg}).  Note
that Eq.~(\ref{eq:cosm})\ contains a potentially large logarithm,  but
at large $\phi$,  $\Omega$, being of  $O(m^4)$, is negligible in any
event. We will explicitly  verify this presently.  So  the behaviour is
controlled by  the UV evolution of the couplings. 

Let us first consider
$Y(0) > Y_{+}$. In that case, as described in the last section both $h$ 
and $\lambda$ increase until $\lambda$ reaches a Landau pole. Before
this the theory of course becomes non-perturbative; the potential is
never lower than it is at the origin while perturbation theory remains
reliable.  If we invoke a UV cut-off $\Lambda$  representing the scale
of new physics then the theory  may or may not remain perturbative up to
the cut-off. In either event, we conclude that it is possible that 
$\phi = 0$ represents the ground state of the theory; we cannot be
certain either because of the non-perturbative nature of the theory in
the UV, or because of the new physics beyond the cut-off.

Now consider the case $Y_{-} < Y_0 < Y_{+}$. Now as $t$ increases $Y$ flows towards 
the fixed point at $Y=Y_{-}$. Thus $\lambda$ and eventually $V$ become negative; 
depending on the initial conditions this can certainly happen for 
perturbative values of the couplings $h,\lambda$.   As mentioned above, just how far one can proceed toward the IR using these $\beta$-functions depends on the value of $m^2,$  since these assume that $m^2$ is sufficiently small so that the scalar particle has not decoupled.

 As an illustration, in Fig.~5 we plot $V(\phi)$ (rather, $V(t)$) as
 calculated from Eq.~(\ref{eq:vrg})
 and from integrating Eq.~(\ref{eq:dvrg})\
 for $n_F = 1$ and values (at $t=0$) $\lambda = h = 1$ and $\mu = m^2$,
 normalising so that
 at $t=0$, $V = 0$. We denote the results for $V$ from the two
 calculations as $V_a$ and $V_b,$ respectively.
The results for $V_a$ and $V_b$ are not precisely the same; the fact 
that they agree well for the range of $t$ displayed is because we are still in the perturbative regime.
We see clearly that at $t=t^* \approx 3.6,$ the potential passes through zero 
becomes rapidly negative. (Note that the potential 
inevitably passes through an extremum (a maximum) before becoming negative; 
we will analyse this extremum in more detail in the special 
case $m^2 = 0,$ which we will consider presently.) 
Evidently we are inclined to deduce that 
the tree minimum at $\phi = 0$ is not the ground state of the theory. 
Can we trust this conclusion? First, note that, with our initial conditions, we 
have $h(t^*) \approx 1.14$ which is manifestly still perturbative.

\begin{figure}
\begin{center}
\includegraphics[angle = -90,totalheight=8cm]{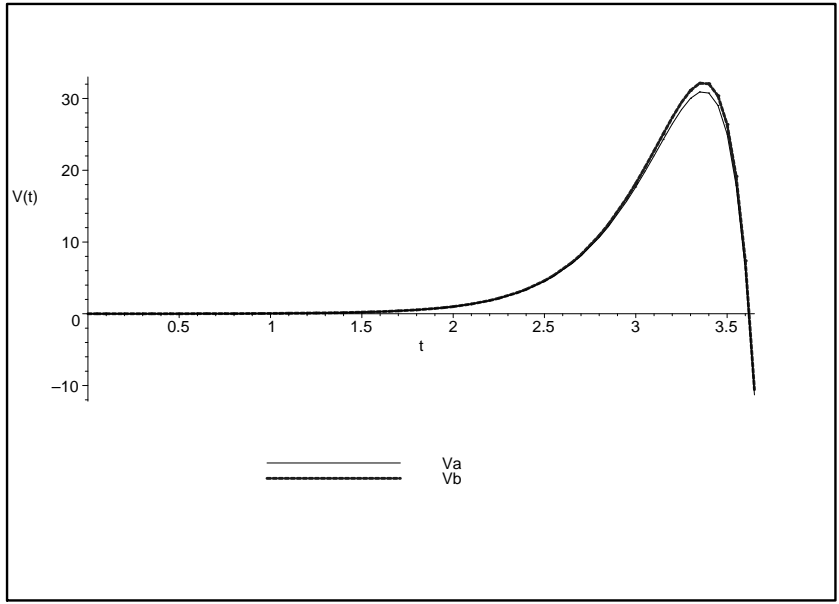}%
\end{center}
\caption{Plot of $V_a$ and $V_b$ against 
$t=\ln \mu$}
\end{figure}

We must also require that the cut-off $\Lambda$  representing the scale
of new physics satisfies $\ln \Lambda >  t^*$, 
if we are to trust the calculation of 
$V(t^*)$. (Of course for values of $\phi$ close to the cut-off 
we would have to consider the effect of 
higher dimensional operators suppressed by one or 
more power of the cut-off~\cite{Datta}-\cite{grzadb}. 
Here we agree with Ref.~\cite{Burgess} that the effect of such 
operators on instability bounds is necessarily small whenever it 
can be reliably calculated.) 
Now for fixed $h(0)$, as we increase $\lambda(0)$ 
(and hence $Y(0)$), $t^*$ increases.   Thus for a given cut-off 
$\Lambda$, the value of $\lambda(0)$ such
that  $e^{t^*} = \Lambda$ is the upper limit on $\lambda(0)$ corresponding 
to a theory such that we can conclude that the potential becomes negative.    
A separate issue is that for sufficiently
large $t$,  $h(t)$ becomes large and we can no longer trust our
calculation. For $e^{t^*} < \Lambda$, therefore,   we cannot 
necessarily deduce that the theory has no ground state; 
but  we claim we can conclude that the
ground state (if it exists)  is lower than the tree minimum.\footnote{In 
Ref.~\cite{BF} it is asserted that after decreasing for  a
certain range of energies, $\lambda(t)$ increases towards a Landau pole;
this is incorrect.  For $Y_{-} \le Y(0) \le Y_{+}$, $\lambda(t)$ decreases
monotonically with $t$ while for $Y(0) > Y_{+}$  $\lambda(t)$ increases
monotonically with $t$.}

Note that $V$ passes through zero at $t\approx 3.6$. With the same 
initial conditions, the coupling $\lambda$ passes through zero at
$t = t^{\prime} \approx 2.97$, and $M^2$ passes through zero at $t\approx 2.98$.
Therefore  for $t > 2.98$ the potential (including one loop corrections)
in fact has  an explicit perturbative imaginary part; in Fig.~5 we have
taken the real part of the potential.  
This imaginary part is similar to that which develops in $ \phi^4$-theory for  
$m^2 < 0$ as $\phi$ is reduced. 
 
Now at  $M^2 < 0$ we have argued above that the imaginary part  
of the potential has a straightforward interpretation. Let us consider the 
potential precisely at the value of $\phi$ defined by $M^2 = 0$. 
Beginning at $L = 4$, the potential contains power 
singularities at $M^2 = 0$, as explained in section~2  
in the context of the pure scalar theory. (Note that in
the $m^2 < 0$ case, $M^2 = 0$ occurs  at a positive value of $\lambda$,
and so happens at both a large value of $\phi$ (as $\lambda$ approaches
zero) and a small value.) Reverting to the $m^2 > 0$ case we have been
discussing,  we must conclude that the perturbation expansion for $V$
breaks down for values of $t$ such that $M^2 \approx 0$.  In Fig.~5
this corresponds to $t \approx 2.98$, as we have already mentioned.
However, at $t^* \approx 3.6$,  where $V$ passes through zero, we find
that $\eta \approx -0.004$; the range of $\phi$ for which $\eta > 1$  
(and hence for which $V$ cannot be reliably calculated in perturbation 
theory) is very small indeed. 

The arguments presented in Ref.~\cite{BF}\ amount essentially to the
statement that  the region of validity of $V(\phi)$ is curtailed by the
requirement that $\lambda(\Lambda) > 0$;  and they propose that the
region of validity of $V(\phi)$ be defined by the inflection point  in
the potential (if this occurs a scale below $\Lambda$),  which we see
from Fig.~5 indeed occurs at the point where $\lambda$ changes  sign as
described above. They further conclude that this  model does not exhibit
the vacuum instability described above;   for $m^2 > 0$ the ground state
of the theory is at $\phi = 0$, and the potential is  a convex function within 
its region of validity (simply because this region is 
curtailed at the inflection point. We simply disagree with this analysis; 
as we have described above, the work of Refs.~\cite{wewu} and~\cite{Guth}
demonstrate quite convincingly, in our opinion, that the potential has 
meaning in its non-convex region; and we see no reason to require that 
$\lambda(\Lambda) > 0$. As usual, we need to require, for perturbative 
believability, that relevant expansion parameters are 
``sufficiently'' small; and in our examples, since we use 
renormalisation group-improved perturbation theory, at large $\phi$ 
these parameters are simply the dimensionless couplings 
evaluated at a scale commensurate with $\phi$ in the UV region. 
It seems clear to 
us, therefore, that for the choice of parameters used  in Fig.~5, 
we can certainly conclude that the local minimum  $\phi = 0$ is unstable. 

It is essentially for the same reason that we disagree with 
Refs.~\cite{kuti},~\cite{holland}. There the claim is that the
calculation of  the potential in a theory without a UV cut-off disagrees
with same calculation  {with\/} a cut-off and that the latter agrees
with simulation data. We claim, however,  that the exhibition of this
disagreement occurs invariably in regions of parameter space  where
perturbation theory is not to be trusted.  When there is substantial
disagreement  between the theory with and without a cutoff, it is a sure
sign that the scale of the field is not far below the scale of the
cutoff.  Perturbation theory breaks down because either $h$ or $\lambda$
become large.  This has, therefore, no bearing on  our discussion above,
where we are always careful 
to confine  our conclusions to 
regions of parameter space where  perturbation theory is expected to
remain valid. 

\subsection{The $m^2 = 0$ case}

We turn now to the interesting case when $m^2 = 0$. The analysis of this 
can be done with the full RG apparatus developed above, but it 
is both interesting and illuminating to mimic the original 
simplified CW 
treatment of scalar QED. Let us suppose for simplicity 
that $0 < Y (0) < Y_{+}$. 
For this range we have seen in section~2 that $\lambda$ approaches zero as $t$ 
increases. For $\lambda(t)$  
sufficiently small, we may neglect the higher-order corrections in 
$\lambda,$ so the effective potential may be written as\footnote{For simplicity, we have chosen a different renormalization convention than in Eq.~\ref{effpot}.} 
\be
V_{eff} = \phi^4 \left[\frac{\lambda}{4!}
-n_F\frac{h^4}{16\pi^2}\ln(\phi^2/v^2)\right],
\label{eq:cwpot}
\ee
where the scale $v$ will be defined presently.  
Then
\be
V'_{eff} = 
\phi^3 \left[\frac{\lambda}{3!}
-n_F\frac{h^4}{8\pi^2}\left(1+2\ln(\phi^2/v^2)\right)\right].
\ee
Evidently  $V'_{eff}$ vanishes for a nonzero value of $\phi$ 
as well as at the origin. If we choose the scale $v$ to be the value of 
$\phi$ at this extremum, then we find it occurs at the 
scale where the running couplings satisfy the relation
\be\label{dimtrans}
\lambda=n_F\frac{3h^4}{4\pi^2},
\ee
 at which point 
\be
V''_{eff}= -n_F\frac{h^4}{2\pi^2}\ v^2<0.
\ee
Thus there is a maximum value defined by dimensional  transmutation,
beyond  which the potential steadily decreases. This
discussion is very similar to the CW one for scalar  QED, except that
there, because the $e^4$ contribution to the effective  potential
differs in sign from the $h^4$ one, the extremum is a minimum,  and
Eq.~(\ref{dimtrans}) above 
is replaced by the equation\footnote{In fact CW 
employed a slightly different subtraction procedure 
from the one implied by Eq.~(\ref{eq:cwpot}), leading to the relation 
$\lambda =  +66\kappa e^4$, but this of course has no effect on the 
physics. This shows, however, that if we define the renormalisation scale 
to be equal to $\vev \phi$ at the minimum, the resulting relation between 
$\lambda$ and $h$ still depends on the details of the subtraction procedure.}
$\lambda =  -9\kappa e^4$. 

Returning to the case $m^2>0,$ it should now be clear that the preceding 
analysis will remain valid to a 
good  approximation so long as $m^2 << v^2$, as in fact
is the case  for the values we chose to produce Fig.~5.  In fact, at the
maximum  of $V$ displayed in Fig.~5, we find that $\lambda \sim -0.2$
while  $h \sim 1.12$.  Note that $\lambda$ and $h^4/(4\pi^2)$  are
indeed of the same order of magnitude at the maximum, as we would
expect.\footnote{The fact that $\lambda$ has the opposite sign  from
that suggested by  Eq.~(\ref{dimtrans}) is for the same reason as we
described above in the CW case, simply a matter of renormalization conventions.}
 
In a derivation based on the RG-improved form of the potential 
(Eq.~(\ref{eq:vrg}) or its derivative Eq.~(\ref{eq:dvrg})) 
Eq.~(\ref{dimtrans}) is replaced by the equation~\cite{yamagishi, ej82} 
\be
4\lambda + \beta_{\lambda} = 0,
\ee
but the essential features of the calculation are unaffected.  We might
note that the Yukawa model is more amenable to perturbative treatment
than scalar QED. In the latter case, while the existence of a local
minimum could be unambiguously demonstrated perturbatively, the behaviour
of the potential both in a neighborhood of the origin as well  as for 
large field values was beyond the realm of perturbation theory.   In our
case, we are able to treat both the origin and  the maximum
perturbatively, with the breakdown of  perturbation theory restricted to
large values of the field.

The only caveat on this discussion is that the coupling $h(v)$  be
sufficiently small so that it is in the perturbative regime.   According
to our discussion in the previous section, there is then always a scale
$\lambda$ at which the relation (\ref{dimtrans}) holds.  To conclude, in
the case $m^2=0,$ the origin is metastable and decays to a region where
eventually the couplings are strong.  Once again, whether the theory has
no ground state or is replaced at  high scales by a modified theory
cannot be determined within this framework.  There is hardly a more
compelling application of radiative corrections than dimensional
transmutation,  yet it is completely missed by the arguments of
ref.~\cite{BF} and overlooked in lattice results,
ref.~\cite{kuti,holland}, whose range of mass parameters does not
include this important region.  


\subsection{The $m^2<0$ case}

Finally, we consider the case $m^2<0.$   At the classical level, the
scalar field behaves as in the case of the $\phi^4$~model discussed in
section~\ref{subsec:phi4}, i.e., the potential has minima at 
$\phi=\pm v,$ where
\be
v^2=\frac{-6m^2}{\lambda}.
\ee
The theory undergoes spontaneous symmetry breaking, and the scalar gets
a mass $m_H$ with  $m_H^2=-2m^2=\lambda v^2/3.$  As a result, 
each fermion also gets a mass $m_f=hv.$  The ratio of the masses is 
\be
\frac{m_H^2}{m_f^2}=\frac{\lambda}{3h^2}=\frac{Y}{3}.
\ee
Radiative corrections to this classical behaviour will shift the values
of the minima and the masses, but so long as perturbation theory
holds,\footnote{For the classical behaviour to be valid, $|m^2|$ must be large compared to the scale $v^2$ associated with dimensional transmutation in the $m^2=0$ case.  Otherwise, the form of the effective potential will be more complicated than assumed here.} this qualitative behaviour is
unchanged. The remainder of the discussion can be inferred from 
the behaviour of the
running couplings.  Generically, $Y\sim O(1).$  The behaviour for large
field values leads us to choose the scale $\mu$ to be $O(\phi)$
to avoid large logs; and   the UV behaviour of $Y(\mu)$ has been discussed
in detail in section~2.  For small values of the 
field, however the potential is not  
given correctly by choosing $\mu\sim\phi.$  
When the field is small compared to the
masses of the scalar and the fermions, they both decouple.  The theory in
the infrared is just that of a set of $n_F$ massive free fermions and 
a massive free scalar, so the
running of the couplings below these masses is not correctly given by a
mass-independent renormalization scheme.

\section{The Standard Model}

The study of the possible \SM\ vacuum instability has a long history and a
 substantial literature. The first comprehensive treatment from an  RG 
perspective was in Ref.~\cite{efjs}; subsequent refinements included 
a particularly lucid discussion in Ref.~\cite{quiros}).  The situation 
in the \SM\ is similar to  the Yukawa model we have considered, in that
the one loop  fermion contribution to $V$ potentially destabilises the
electro-weak vacuum; however  it differs in that onset of the
instability is associated with another, deeper,  minimum of the
potential (which may still occur within the perturbative regime).  

The deeper minimum can occur because while the top 
quark Yukawa coupling contribution to the evolution of the Higgs 
self-coupling $\lambda$
evolution tends to push it down towards (or through) zero in the UV, 
the Yukawa coupling itself gets smaller in the UV 
due to the gauge 
coupling contributions to its evolution.\footnote{In the \SM, the 
top Yukawa is substantially below 
the Quasi-Infra-Red Fixed Point value~\cite{hill}
which would correspond to approaching a Landau pole at gauge unification.}  
Therefore eventually the gauge 
coupling contributions to the evolution of $\lambda$ 
cause it to recover to positive values. It is easy to verify this from the 
explicit expressions for the $\beta$-functions~\cite{fjj}, which 
for convenience we reproduce below. (Note that the 
scalar anomalous dimension is gauge dependent: the result below is for the 
Landau gauge.) 

The one-loop RG functions are 

\bea
\ikap\gamma^{(1)}& = &  3h^2-\frak{9}{4}g^2-\frak{3}{4}{g^\prime}^2\nn
\ikap\beta_{\lambda}^{(1)}& = & 4\lambda^2+12\lambda h^2-36h^4-9\lambda g^2
-3\lambda {g^\prime}^2\nn
&+& \frak{9}{4}{g^\prime}^4+\frak{9}{2}g^2{g^\prime}^2+\frak{27}{4}g^4\nn
\ikap\beta_h^{(1)}& = & \frak{9}{2}h^3-8g_3^2 h-\frak{9}{4}g^2 h-\frak{17}{12}
{g^\prime}^2h\nn
\ikap\beta_g^{(1)}& = & -\frak{19}{6}g^3\nn
\ikap\beta_{g^\prime}^{(1)}& = & \frak{41}{6}{g^\prime}^3\nn
\ikap\beta_{g_3}^{(1)}& = & -7g_3^3\nn
\ikap\beta_{m^2}^{(1)}& = & m^2(2\lambda+6h^2-\frak{9}{2}g^2
-\frak{3}{2}{g^\prime}^2).
\eea
The two-loop contributions to the \RG\ functions are given by~\cite{fjj}
\bea
\iikap\gamma^{(2)}& = & \frak{1}{6}\lambda^2-\frak{27}{4}h^4+20g_3^2h^2
+\frak{45}{8}g^2h^2+\frak{85}{24}{g^\prime}^2h^2\nn
&-& \frak{271}{32}g^4+\frak{9}{16}g^2{g^\prime}^2+\frak{431}{96}
{g^\prime}^4\nn
\iikap\beta_{\lambda}^{(2)}& = & -\frak{26}{3}\lambda^3-24\lambda^2h^2
+6\lambda^2(3g^2+{g^\prime}^2)-3\lambda h^4+80\lambda g_3^2h^2\nn
&+& \frak{45}{2}\lambda g^2h^2+\frak{85}{6}\lambda{g^\prime}^2h^2
-\frak{73}{8}\lambda g^4
+\frak{39}{4}\lambda g^2{g^\prime}^2+\frak{629}{24}\lambda{g^\prime}^4\nn
&+& 180h^6-192h^4g_3^2-16h^4{g^\prime}^2-\frak{27}{2}h^2g^4
+63h^2g^2{g^\prime}^2\nn
&-& \frak{57}{2}h^2{g^\prime}^4+\frak{915}{8}g^6-\frak{289}{8}g^4{g^\prime}^2
-\frak{559}{8}g^2{g^\prime}^4-\frak{379}{8}{g^\prime}^6\nn
\iikap\beta_h^{(2)}& = & h\bigl(-12h^4+h^2(\frak{131}{16}{g^\prime}^2
+\frak{225}{16}g^2+36g_3^2-2\lambda)+\frak{1187}{216}{g^\prime}^4\nn
&-& \frak{3}{4}g^2{g^\prime}^2+\frak{19}{9}{g^\prime}^2g_3^2-\frak{23}{4}g^4
+9g^2g_3^2-108g_3^4+\frak{1}{6}\lambda^2\bigr)\nn
\iikap\beta_g^{(2)}& = & g^3(\frak{3}{2}{g^\prime}^2+\frak{35}{6}g^2+12g_3^2
-\frak{3}{2}h^2)\nn
\iikap\beta_{g^\prime}^{(2)}& = & {g^\prime}^3(\frak{199}{18}{g^\prime}^2
+\frak{9}{2}g^2+\frak{44}{3}g_3^2-\frak{17}{6}h^2)\nn
\iikap\beta_{g_3}^{(2)}& = & g^3_3(\frak{11}{6}{g^\prime}^2
+\frak{9}{2}g^2-26g_3^2-2h^2)\nn
\iikap\beta_{m^2}^{(2)}& = & 2m^2\bigl(-\frak{5}{6}\lambda^2-6\lambda h^2
+2\lambda(3g^2+{g^\prime}^2)-\frak{27}{4}h^4+20g_3^2h^2\nn
&+& \frak{45}{8}g^2h^2+\frak{85}{24}{g^\prime}^2h^2-\frak{145}{32}
g^4+\frak{15}{16}g^2{g^\prime}^2+\frak{157}{96}{g^\prime}^4\bigl).
\eea
Here $h,\lambda$ are the Yukawa and Higgs self-coupling normalised so that 
$m_t = hv/\sqrt{2}$ and $\lambda = 6m^2/v^2 = 3m_h^2/m^2$ where $v\sim 246$ GeV.

One easily sees that for a given $h$, there will be a {\it lower\/} limit 
on $\lambda$ such that the electroweak vacuum is the true one; this 
corresponds (for a given $m_t$) to a lower bound on the Higgs mass $m_h$. 

For $m_h$ below this bound the electroweak vacuum would be metastable. 
Here we have little to add to the discussion of Ref.~\cite{quiros}. 
For a cut-off $\Lambda = 1~\TeV,$ the bound is 
about~\cite{quiros}  
\be
m_H (\GeV) > 52 + 0.64 (m_t (\GeV)  - 175) 
- 0.5 \frac{\alpha(M_Z) - 0.118}{0.006},
\ee
leaving us with comforting 
stability,\footnote{For the issue of whether the electroweak 
vacuum might be unstable and yet sufficiently long-lived to permit our 
presence, we refer the reader to the literature; for 
a recent example see Ref.~\cite{alta}.} 
given existing experimental knowledge 
of $m_t$ and bounds on $m_h$. For $\Lambda = 10^{19}~\GeV$, the bound 
$m_H (\GeV) > 52~\GeV$ above becomes $m_H (\GeV) > 134~\GeV$, 
so it remains possible that the discovery of a 
Higgs (with $m_H (\GeV) <  134~\GeV$) 
will provide, from this point of view, evidence for 
physics beyond the standard model. 

The $M^2 = 0$  instability on which we remarked occurs in this case too, and
is similarly confined to  a very narrow range of $\phi$ values well away
from the value of $\phi$ at which  the potential drops below the
electro-weak minimum.  

The SM analysis is repeated in Ref.~\cite{BF}\
where their philosophy that the physical cut-off may not be  higher than
the inflection point in the potential leads, (for smaller values of the
cut-off)  to substantially different lower bounds on $m_H$ than those
obtained in Ref.~\cite{quiros}. They also assert that  the metastable
scenario described above cannot occur, due essentially to the convex 
nature of the potential.  We maintain that is not correct, as we have explained above. 

\section{Conclusions}

Using mainly the simple $\lambda\phi^4$ theory and its extension 
involving Yukawa couplings, we have reviewed aspects of the 
computation of the scalar effective potential that are sensitive to 
non-convex regions of the potential, with a particular focus on the issue 
of ground state stability.  We have found no reason to disbelieve that 
a sufficiently large fermionic contribution will destabilise the tree vacuum (corresponding 
to the origin in a simple theory with scalar $m^2 \ge 0$ or to the electroweak 
vacuum in the \SM.)  Moreover this instability 
can be consistently demonstrated within the confines of perturbative 
believability.  The Yukawa model we have studied has the 
additional interesting feature that it exhibits weak-coupling
dimensional transmutation associated with a local {\it maximum\/} of the 
potential.  The scale associated with dimensional transmutation is crucial, 
since it sets the value of the field where the instability may occur, even when $m^2 \ne 0.$

We conclude that calculations of the effective potential within the
perturbative domain can be trusted, and inferences about metastability
relied upon.  While the ultimate effective potential may be convex, or
the constrained effective potential rather different from the
perturbative calculation, their application are not in contradiction
with this conclusion.  One simply must be careful about the questions
one wishes to address.  The formalism employed must conform to the
particular physical situation under discussion.

\section*{\large Acknowledgements}

DRTJ was visiting KITP (Santa Barbara) while part of this work was done. This
work was partially supported by the National Science Foundation
under Grant No. PHY99-07949.


%
%
%
%
%
\end{document}